# Electrical Control of Carriers' Spin Orientation in FeVTiSi Heusler Alloy

Jiahui Zhang,[1] Xingxing Li,[1] and Jinlong Yang[1,2*]

[1]*Hefei National Laboratory for Physical Science at the Microscale, University of Science and Technology of China, Hefei, Anhui 230026, China.*

[2]*Synergetic Innovation Center of Quantum Information and Quantum Physics, University of Science and Technology of China, Hefei, Anhui 230026, China.*

The direct control of carrier's spin by electric field under room temperature is one of the most important challenges in the field of spintronics. For this purpose, we here propose a quaternary Heusler alloy FeVTiSi. Based on first principles calculations, FeVTiSi alloy is found to be an intrinsic bipolar magnetic semiconductor in which the valence band and conduction band approach the Fermi level through opposite spin channels. Thus FeVTiSi alloy can conduct completely spin-polarized currents with tunable spin-polarization direction simply by applying a gate voltage. Furthermore, by Monte Carlo simulations based on the classical Heisenberg Hamiltonian, the Curie temperature of FeVTiSi alloy is predicted to be as high as 1293 K, far above the room temperature. The bipolar magnetic semiconducting character and the high Curie temperature endow the FeVTiSi alloy great potentials in developing electrically controllable spintronic devices working at room temperature.

PACS numbers:

Spintronic materials [1-7] have attracted much attention in the past decades due to their unique merits in data storage, transportation and processing. Half metals (HMs) are an important class of spintronic materials which can provide completely spin-polarized currents [8,9]. The carrier's spin orientation of HMs is however fixed at one direction and an external magnetic field is needed to switch between two spin-polarization directions. Nevertheless, it is more convenient to use electric fields rather than magnetic fields since electric fields can be easily applied locally.

To solve this problem, several schemes have been proposed from both experimental and theoretical aspects. For example, the electrical control of spin orientation can be achieved experimentally in a semiconducting structure with spatially varying g-tensor [10], or simply via the effect of spin-orbital coupling [11]. The magneto-electric effect in multiferroic systems also provides a feasible route to control of magnetism by electric field [12]. Theoretically, a new class of spintronic materials, bipolar magnetic semiconductors (BMS), were also proposed for this purpose [13,14]. In BMS, the valence bands (VB) and conduction bands (CB) are fully spin-polarized in the opposite spin directions, i.e. there exist up-spin electrons in VB and down-spin electrons in CB. By applying a gate voltage, the electrons in VB or CB can be selected for conduction through changing the gate polarity, therefore completely spin-polarized currents with gate-tunable spin direction can be obtained. This is totally different from the above mentioned schemes, which are focused on tuning the spatial orientation of specific spins. Moreover, the control of spin in BMS does not rely on any varying g-tensor, spin-orbital coupling or magneto-electric effect, and is more convenient to implement. Though some BMS materials have been found [13-25], their applications are hindered either by the difficulty in experimental preparation or the low Curie temperature. It is still a big challenge to find BMS materials available at room temperature.

In this paper, we propose an experimentally accessible bipolar magnetic semiconductor with high Curie temperature, the FeVTiSi alloy. This compound belongs to the big family of quaternary Heusler alloys, which can be identified by the formula XX'YZ, where X, X' and Y are transition metals while Z is a main-group element. In experiment, quaternary Heusler alloys can be prepared by repeated arc melting of stoichiometric mixtures of high-purity elements in an inert atmosphere [26,27]. By altering the four different composite elements, quaternary Heusler alloys provide a fertile soil for exploring various spintronic materials such as HMs and spin gapless semiconductors [28-33]. The successes in previous works inspire us to find bipolar magnetic semiconductors among such alloys.

Based on first principles calculations, a quaternary Heusler alloy FeVTiSi is designed. The electronic structure study shows that it is a typical BMS material with a magnetic moment of 3.00 $\mu_B$ per unit cell. Monte Carlo simulation is performed to investigate the magnetic stability of the alloy, and the Curie temperature is predicted as high as 1293 K. Thus FeVTiSi alloy is a good candidate for application in electrical control of carriers' spin orientation at room temperature.

Geometry optimization and magnetic studies were performed within the Perdew-Burke-Ernzerhof generalized gradient approximation (GGA) [34] with strong-correlated correction using the method of GGA+U [35] implemented in the Vienna ab initio Simulation Package (VASP) [36]. The effective onsite Coulomb interaction parameter $U_{eff}$ of Fe, V and Ti was set to be 1.80, 1.34 and 1.36 eV respectively, followed by a previous work on similar systems [37]. A supercell that contains eight FeVTiSi units was used and the Brillouin zone was sampled by a $3\times6\times6$ Monkhorst-Pack k-points mesh [38]. The energy cut off was set to 500 eV. The convergence of energy and force was set to $1\times10^{-5}$ eV and 0.01 eV/Å, respectively. Based on the magnetic exchange parameters calculated with GGA+U, we performed Monte Carlo simulations [39] with a $12\times12\times12$ supercell to estimate the Curie temperature of FeVTiSi alloy. Then the electronic properties were computed by employing the screened hybrid HSE06 functional [40,41], which usually performs much better than the GGA and GGA+U methods. Herein, a primitive cell containing only one FeVTiSi unit was used with an energy cut off of 400 eV. The Brillouin zone was sampled by a set of $7\times7\times7$ k-points. The effect of spin-orbital coupling (SOC) is not considered since our test calculations show that it has little influence on our results.

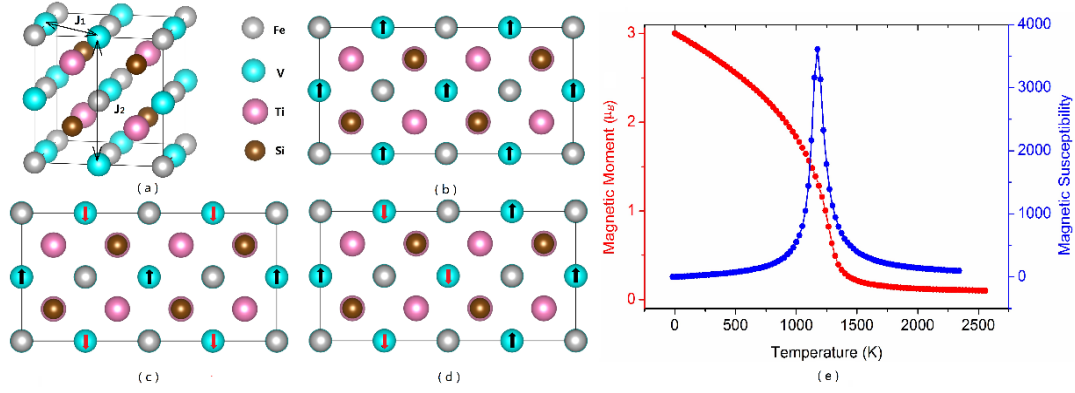

FIG. 1 (color online). (a) Optimized geometry of FeVTiSi alloy. $J_1$ and $J_2$ indicate the nearest and next nearest exchange parameter, respectively. Silver, blue, pink and brown balls indicate ferrum, vanadium, titanium and silicon atoms respectively. Illustration of (b) FM, (c) AFM and (d) AFM2 states along (100) direction. (e) Variation of magnetic moment per unit cell (red lines) and magnetic susceptibility (blue lines) with respect to temperature calculated by the Monte Carlo simu-lations based on the classical Heisenberg Hamiltonian.

In a typical structure of quaternary Heusler alloy XX'YZ, X atoms occupy the positions of (0,0,0), (0,0.5,0.5), (0.5,0.5,0) and (0.5,0,0.5) in fraction coordinate; X' atoms occupy the position of (0,0,0.5), (0,0.5,0), (0.5,0,0) and (0.5,0.5,0.5); Y atoms occupy the positions of (0.25,0.25,0.25), (0,75,0.75,0.25), (0.75,0.25,0.75) and (0.25,0.75,0.75); Z atoms occupy the position of (0.25,0.75,0.25), (0.75,0.25,0.25), (0.25,0.25,0.75) and (0.75,0.75,0.75). For FeVTiSi alloy, we have X= Fe, X'= V, Y= Ti and Z= Si (Fig. 1a). The optimized lattice parameter is 5.90 Å. We have also tested two competitive structures, FeTiVSi and VTiFeSi, and find they are much less stable than FeVTiSi by 0.669 eV and 0.517 eV per unit cell in total energy, respectively. Thus our proposed structure is the most stable one.

It is well known Heusler alloys satisfy the Slater-Pauling rule [42,43], from which the magnetic moment M per unit cell is decided by the following equation:

$$M = |N_v - 24| \quad (1)$$

where $N_v$ is the total number of valence electrons per unit cell. For FeVTiSi, $N_v$= 21. Thus we anticipate that FeVTiSi alloy is a ferromagnet bearing a magnetic moment of

3 μB per unit cell, which has been verified by our first principles calculations. Further study shows that the magnetic moment is mainly contributed by V element with 2.95 μB per atom, while Fe and Ti carry minor moments with equal value, but in the opposite direction.

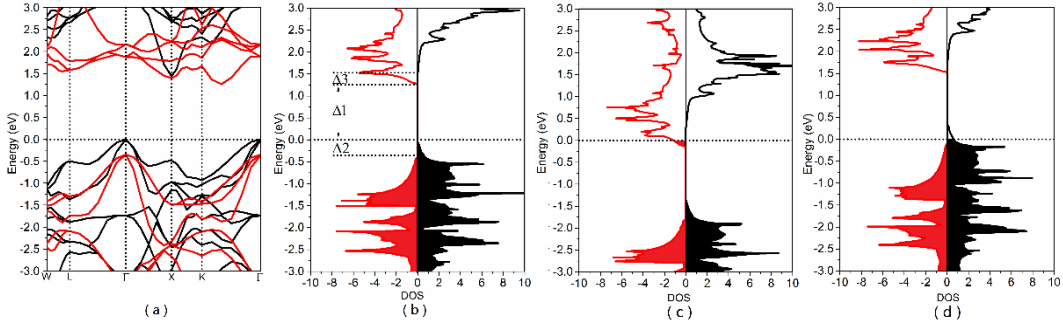

FIG. 2 (color online). Calculated (a) band structure and (b) DOS of FeVTiSi based on HSE06 functional. The DOS under (c) electron doping and (d) hole doping with a doping concentration of 0.025 carrier per atom. Black and red indicate spin up and spin down, respectively. The Fermi level is set at zero.

To estimate the Curie temperature of FeVTiSi alloy, we perform Monte Carlo simulations based on the classical Heisenberg Hamiltonian

$$H = -\sum_{i,j} J_1 S_i S_j - \sum_{k,l} J_2 S_k S_l \quad (2)$$

where $J_1$ and $J_2$ are the nearest and next-nearest exchange parameters respectively, and S= 3/2 is the effective spin of V atoms. Here, we simplify our model by considering that only V atoms carry spins. To determine $J_1$ and $J_2$, we calculate the total energy of three different magnetic states: ferromagnetic (FM) state (Fig. 1b), antiferromagnetic (AFM) state (Fig. 1c) and double-layer antiferromagnetic (AFM2) state (Fig. 1d). The results show FM state is 1.69 and 1.05 eV lower in energy than AFM and AFM2 states, respectively. From the two energy differences, we deduce $J_1$= 11.732 meV and $J_2$= 5.688 meV according to the following equations.

$$E_{FM} = -48J_1 S^2 - 24J_2 S^2 + E_0 \quad (3)$$

$$E_{AFM} = 16J_1 S^2 - 24J_2 S^2 + E_0 \quad (4)$$

$$E_{AFM2} = -16J_1 S^2 - 8J_2 S^2 + E_0 \quad (5)$$

$$J_1 = (E_{AFM} - E_{FM})/64S^2 \quad (6)$$

$$J_2 = [32J_1S^2 - (E_{AFM} - E_{AFM2})]/16S^2 \qquad (7)$$

The simulated spin magnetic moment and susceptibility as a function of temperature are plotted in Fig. 1e, from which one can see that the spin magnetic moment does not fall to a negligible value until about 1300 K and the Curie temperature of FeVTiSi is predicted to be 1293 K by locating the peak position in the plot of magnetic susceptibility. Considering that this method usually gives a good estimate for the experimental Curie temperature [39,44], we expect FeVTiSi alloy keeps its ferromagnetism at room temperature.

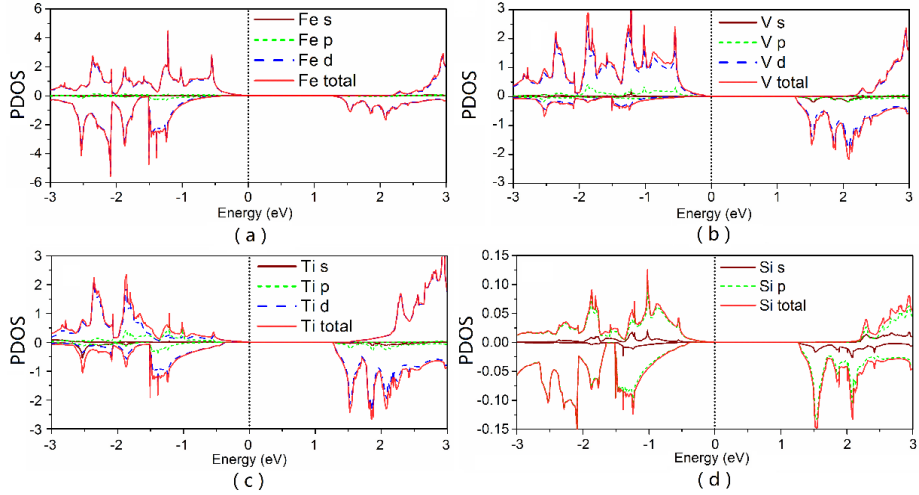

FIG. 3 (color online). Orbital-projected density states (PDOS) of (a) Fe element, (b) V element, (c) Ti element and (d) Si element calculated by HSE06 functional. The Fermi level is set at zero.

Electronic structure calculations show that FeVTiSi alloy is a ferromagnetic semiconductor with an indirect gap of 1.26 eV (Fig. 2a). Interestingly, we find inverse spin-polarization directions for valence band and conduction band, that is, VB and CB approach the Fermi level through opposite spin channels. Therefore, FeVTiSi alloy is a BMS. A typical BMS can be characterized by three energy gaps $\Delta 1$, $\Delta 2$ and $\Delta 3$ as shown in Fig. 2b. $\Delta 1$ represents the spin-flip gap between VB and CB edges from different spin channels. $\Delta 1+\Delta 2$ and $\Delta 1+\Delta 3$ reflect the spin-conserved gaps for two spin channels, respectively [13]. For FeVTiSi alloy, we predict $\Delta 1$= 1.26 eV, $\Delta 2$= 0.36 eV and $\Delta 3$= 0.27 eV.

Doping holes or electrons into the system will shift the Fermi level into VB or CB. Since VB and CB are fully spin polarized, it turns into half metals with different spin polarization directions depending on the doping type. For example, under electron doping, FeVTiSi alloy changes to a half metal with spin-down polarization (Fig. 2c), while a spin-up polarized half metal is obtained under hole doping (Fig. 2d). Thus completely spin-polarized currents with tunable spin-polarization can be realized simply by applying a voltage gate upon the FeVTiSi alloy. Note that in bulk FeVTiSi, the possible existent defects and traps may make the Fermi level pinned and difficult to tune. This problem can be solved by using FeVTiSi in its thin film form, since large scale Fermi level shift can be easily achieved in two dimensional systems via a gate voltage [45-47].

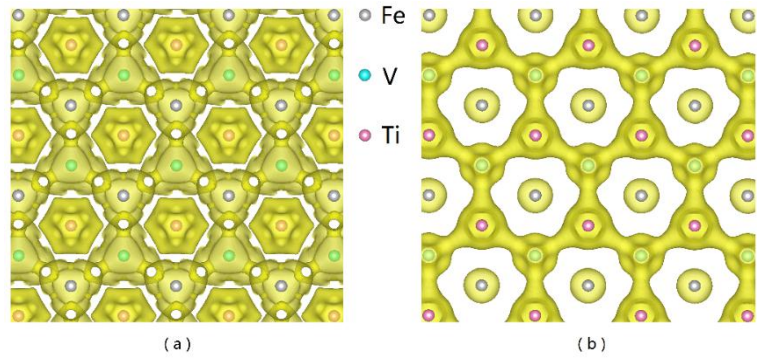

FIG. 4 (color online). Isosurfaces of (a) valence band charge in the energy window of $\Delta 2$ and (b) conduction band charge in the energy window of $\Delta 3$ viewed along (111) direction. The isovalues are set to be 0.0003 e/Å$^3$ and 0.003 e/Å$^3$, respectively.

The orbital-projected density of states (PDOS) of each element are further calculated to understand the electronic properties of FeVTiSi alloy. From Fig. 3, it can be seen that the states around the Fermi level are mainly contributed by the 3d orbitals of Fe, V and Ti elements. Though only V element carries the most magnetic moment of the system, all the PDOS of the four elements are polarized and show BMS character in different degrees. The Fe and V elements contribute to the main spin-up states below the Fermi level while V and Ti elements construct the most spin-down states above the Fermi level.

Finally, we plot the charge density of VB and CB in the energy window of Δ2 and Δ3 respectively (Fig. 4), in which we set the isovalue to be 0.0003 e/Å$^3$ for Δ2 while 0.003 e/Å$^3$ for Δ3. This is because the charge density in Δ2 is about ten times lower than that in Δ3, which can be also seen from Fig. 2b where there is a peak of DOS in Δ3 while none in Δ2.

Interestingly, the spatial distributions of the charge density of VB and CB are different. For VB, the charge density of Fe and V atoms are delocalized and connected together while that of Ti atoms are mainly localized near their atom sites. For CB, the charge density of V and Ti atoms are delocalized and linked together but that of Fe atoms are mainly concentrated around the Fe atoms. This indicates that in the case of hole doping, the spin up electrons can travel along Fe-V-Fe route, while in the case of electron doping, the spin down electrons are transported along the path of V-Ti-V.

In conclusion, based on first principles calculations, we have proposed a ferromagnetic quaternary Heusler alloy FeVTiSi with intrinsic bipolar magnetic semiconducting character. Monte Carlo simulations suggest the FeVTiSi alloy possess a rather high Curie temperature of 1293 K. In FeVTiSi alloy, VB and CB edges are completely spin-polarized in the opposite spin direction. Thus electron and hole doping result in half-metallic conduction in different spin channels, which can be easily realized by voltage gate. Moreover, both the charge density of VB and CB are partly delocalized, ensuring a good conducting behavior under carrier doping. Therefore, the FeVTiSi alloy is promising for application in room-temperature electrical control of carriers' spin orientation.

This work is partially supported by the National Key Basic Research Program (2011CB921404, 2012CB922001), by NSFC (21121003, 91021004, 20933006, 11004180, 51172223), by Strategic Priority Research Program of CAS

*Author to whom correspondence should be addressed. Electronic mail: jlyang@ustc.edu.cn.

**References**

1. M. N. Baibich, J. M. Broto, A. Fert, F. Nguyen Van Dau, F. Petroff, P. Etienne, G. Greuzet, A. Friederich, and J. Chazelas, Phys. Rev. Lett. **61**, 2472 (1988).
2. G. P. Binash, P. Grünberg, F. Saurenbach, and W. Zinn, Phys. Rev. B **39**, 4828 (1989).
3. S. A. Wolf, D. D. Awschalom, R. A. Buhrman, J. M. Daughton, S. von Molná, M. L. Roukes, A. Y. Chtchelkanova, and D. M. Treger, Science **294**, 1488 (2001).
4. I. Žutić, J. Fabian, and S. Das Sarma, Rev. Mod. Phys. **76**, 323 (2004).
5. D. D. Awschalom and M. E. Flatté, Nat. Phys. **3**, 153 (2007).
6. J. M. Zhang, D. Gao, and K. W. Xu, Sci. China Phys. Mech. Astron. **55**, 428 (2012).
7. P. Chen and G. Y. Zhang, Sci. China Phys. Mech. Astron. **56**, 207 (2013).
8. R. A. de Groot, F. M. Mueller, R. G. van Engen, and K. H. J. Buschow, Phys. Rev. Lett. **50**, 2024 (1983).
9. H. van Leuken and R. A. de Groot, Phys. Rev. Lett. **74**, 1171 (1995).
10. Y. Kato, R. C. Myers, D. C. Driscoll, A. C. Gossard, J. Levy and D. D. Awschalom, Science **299**, 1201 (2003).
11. K. C. Nowack, H. F. L. Koppens and Y. V. Nazarov, Science **318**, 1430 (2007).
12. S.-W. Yang, R.-C. Peng, T. Jiang, Y.-K. Liu, L. Feng, J.-J. Wang, L.-Q. Chen, X.-G. Li, and C.-W. Nan, Adv. Mater. 2014, DOI: 10.1002/adma.201402774.
13. X. X. Li, X. J. Wu, Z. Y. Li, J. L. Yang, and J. G. Hou, Nanoscale **4**, 5680 (2012).
14. X. Li and J. Yang, Phys. Chem. Chem. Phys. **15**, 15793 (2013).

15. P. Mahadevan and A. Zunger, Phys. Rev. B **69**, 115211 (2004).

16. S. Sanvito, P. Ordejón, and N. A. Hill, Phys. Rev. B **63**, 165206 (2001).

17. I. I. Mazin and D. J. Singh, Phys. Rev. B **56**, 2556 (1997).

18. L. Yuan, Z. Li, and J. Yang, Phys. Chem. Chem. Phys. **15**, 497 (2013).

19. X. Li, X. Wu, and J. Yang, J. Mater. Chem. C **1**, 7197 (2013).

20. J. Zhang, X. Li, and J. Yang, Appl. Phys. Lett. **104**, 172403 (2014).

21. J. Zhou, Q. Wang, Q. Sun, X. S. Chen, T. Kawazoe, and P. Jena, Nano lett. **9**, 3867 (2009).

22. Y. Ding and Y. Wang, Appl. Phys. Lett. **102**, 142115 (2013).

23. Y. Ding and Y. Wang, Appl. Phys. Lett. **104**, 083111 (2014).

24. J. Li, Z. H. Zhang, D. Wang, Z. Zhu, Z. Q. Fan, G. P. Tang, and X. Q. Deng, Carbon **69**, 143 (2013).

25. X. Li, X. Wu, and J. Yang, J. Am. Chem. Soc. **136**, 11065 (2014).

26. V. Alijani, J. Winterlik, G. H. Fecher, S. S. Naghavi, and C. Felser, Phys. Rev. B **83**, 184428 (2011).

27. Xu. Dai, G. Liu, G. H. Fecher, C. Felser, Y. Li, and H. Liu, J. Appl. Phys. **105**, 07E901 (2009).

28. X. L. Wang, Phys. Rev. Lett. **100**, 156404 (2008).

29. S. Izadi and Z. Nourbakhsh, J. Supercond. Nov. Magn. **24**, 825 (2011).

30. G. Gökoğlu, Solid State Sci. **14**, 1273 (2012).

31. G. Z. Xu, E. K. Liu, Y. Du, G. J. Li, G. D. Liu, W. H. Wang, and G. H. Wu, Europhys. Lett. **102**, 17007 (2013).

32. G. Y. Gao, Lei Hu, K. L. Yao, Bo Luo, and Na Liu, J. Alloys Compd. **551**, 539 (2013).

33. I. Galanakis, K. Özdoğan, and E. Şaşıoğlu, Appl. Phys. Lett. **103**, 142404 (2013).

34. G. Kresse and J. Furthmüller, Phys. Rev. B: Condens. Matter. **54**, 11169 (1996).

35. J. P. Perdew, K. Burke, and M. Ernzerhof, Phys. Rev. Lett. **77**, 3865 (1996).

36. H. J. Monkhorst and J. D. Pack, Phys. Rev. B **13**, 5188 (1976).

37. A. I. Liechtenstein and V. I. Zaane, J. Phys. Rev. B **52**, 5467 (1995).

38. H. C. Kandpal, G. H. Fecher, and C. Felser, J. Phys. D: Appl. Phys. **40**, 1507

(2007).

39. H. J. Xiang, S. H. Wei, and M. H. Wangbo, Phys. Rev. Lett. **100**, 167207 (2008).
40. J. Heyd, G. E. Scuseria, and M. Ernzerhof, J. Chem. Phys. **118**, 8207 (2003).
41. J. Heyd, G. E. Scuseria, and M. Ernzerhof, J. Chem. Phys. **124**, 219906 (2006).
42. T. Graf, C. Felser, and S. S. P. Parkin, Prog. Solid State Chem. **39**, 1 (2011).
43. K. Özdoğan, E. Şaşıoğlu, and I. Galanakis, J. Appl. Phys. **113**, 193903 (2013).
44. X. Li and J. Yang, J. Mater. Chem. C **2**, 7071 (2014).
45. Y.-W. Tan, Y. Zhang, K. Bolotin, Y. Zhao, S. Adam, E. H. Hwang, S. Das Sarma, H. L. Stormer and P. Kim, Phys. Rev. Lett. **99**, 246803 (2007).
46. H. Yuan, H. Shimotani, A. Tsukazaki, A. Ohtomo, M. Kawasaki, and Y. Iwasa, Adv. Funct. Mater. **19**, 1046 (2009).
47. A. S. Dhoot, C. Israel, X. Moya, N. D. Mathur, and R. H. Friend, Phys. Rev. Lett. **102**, 136402 (2009).